\documentclass{optica-article}

\journal{opticajournal} 

\articletype{Research Article}

\usepackage{lineno}

\begin{document}

\title{The Bell-Bloom-type optically-pumped FID Rubidium atomic magnetometer with a multi-passing probe beam and two counter-propagating pump beams}

\author{Yongbiao YANG,\authormark{1} Zhengyu SU,\authormark{1} Yang LI,\authormark{1} Yanhua WANG,\authormark{2,3},Jun HE\authormark{1,3},Xiaojun JIA\authormark{1,3} and Junmin WANG\authormark{1,3,*}}

\address{\authormark{1}State Key Laboratory of Quantum Optics Technologies and Devices, and Institute of Optoelectronics, Shanxi University, Taiyuan 030006, China\\
\authormark{2}School of Physics and Electronic Engineering, Shanxi University, Taiyuan 030006, China\\
\authormark{3}Collaborative Innovation Center of Extreme Optics, Shanxi University, Taiyuan 030006, China}

\email{\authormark{*}wwjjmm@sxu.edu.cn} 


\begin{abstract*} 
The Bell-Bloom-type optically-pumped atomic magnetometers are well-suited for weak geomagnetic field detection. However, conventional single-beam pumping introduces an atomic spin polarization gradient, which limits the measurement accuracy and sensitivity. To address this issue, this paper proposes and experimentally demonstrates a Bell-Bloom-type rubidium FID magnetometer scheme integrating orthogonally polarized counter-propagating pumping and multi-pass probe detection. This design homogenizes the atomic spin polarization distribution, suppresses light shifts and power broadening effects induced by the pump beam, Meanwhile, the five-pass probe configuration significantly enhances the signal amplitude. Experimental results reveal that, compared with the traditional single-beam pumping and single-pass detection scheme, the proposed magnetometer achieves a remarkable improvement in magnetic field measurement accuracy, and the magnetic field sensitivity is improved from 18.9 pT/$\sqrt{\text{Hz}}$ to 3.1 pT/$\sqrt{\text{Hz}}$. This work provides an effective technical approach and reference for  optimizing the performance of atomic magnetometers and extending their applications in integrated arrays.

\end{abstract*}

\section{Introduction}
In the field of precise magnetic field measurement, various weak magnetic field sensors have been developed based on different working principles, such as cavity optomechanical magnetometers \cite{li2020ultrabroadband}, NV center magnetometers \cite{fang2013high,clevenson2015broadband}, and atomic magnetometers \cite{sander2012magnetoencephalography,li2022kilohertz}. These sensors have been widely applied in numerous fields, including magnetoencephalography and magnetocardiography \cite{zhang2020recording}, geological exploration \cite{nabighian2005historical}, materials science \cite{kim2017magnetic}, dark matter search \cite{jiang2021search}, and fundamental physics research \cite{kim2018experimental}. Among various weak magnetic field sensors, atomic magnetometers have become one of the most promising due to their high sensitivity, wide measurement range, and potential for miniaturization and on-chip integration. However, existing atomic magnetometers are still limited by issues such as light-induced frequency shifts and spin polarization gradients. Particularly for Bell-Bloom magnetometers operating in non-SERF states, these bottlenecks have not yet been effectively resolved.

 Atomic magnetometers operate based on the principle of atomic spin precession in an external magnetic field and can be classified into various categories according to their work modes. Spin-Exchange Relaxation-Free (SERF) magnetometers have replaced Superconducting Quantum Interference Devices (SQUIDs) due to their ultrahigh sensitivity and are highly favored in zero-magnetic-field environments \cite{li2018serf}. For geomagnetic field measurements, radio-frequency (RF)-driven and synchronous-pumping schemes have become mainstream techniques. Compared with RF-driven spin polarization, the synchronous-pumping approach offers notable advantages in terms of low crosstalk thanks to its all-optical configuration. This optical technique for atomic spin polarization was first experimentally demonstrated by Bell and Bloom in 1961 \cite{bell1961optically}. Its core mechanism is to precisely synchronize the modulation frequency of the pump light with the Larmor precession frequency of atoms in the magnetic field under measurement. The corresponding all-optical magnetometer is called the Bell-Bloom optically pumped magnetometer, which has been extensively studied in recent decades \cite{higbie2006robust,qin2024enhanced}. 

High atomic number density induces strong resonant absorption and rapid attenuation of circularly polarized pump light. As a result, atoms near the pump light input absorb most of the optical energy, while atoms farther downstream receive insufficient optical power. This reduces the overall level of atomic polarization and introduces spatial inhomogeneity in atomic spin polarization. Although the spin polarization gradient issue in SERF magnetometers has been widely studied with multiple solutions proposed \cite{sun2025polarization,zhao2020improvement,peng2024femtotesla}, research on this problem in non-SERF magnetometers remains scarce. Even though non-SERF magnetometers do not require ultrahigh atomic number density, pump light attenuation still induces a notable spin polarization gradient, which has become a critical bottleneck hindering further performance improvement of such devices. In addition, the limited light-atom interaction length in traditional single-pass probe detection results in weak detected signals and a low signal-to-noise ratio (SNR), further restricting the sensitivity enhancement of  magnetometers.

The strong attenuation of pump light at high atomic number density also induces optical frequency shifts\cite{zou2026modeling}, which typically limit the accuracy of atomic magnetometers but can be effectively suppressed by operating the devices in pulsed mode. After optical pumping, the atomic spin precession exhibits free induction decay (FID) when the pump beam is switched off. Detecting the FID signal for atomic spin also serves as a standard method to characterize the transverse spin relaxation time ($T_2$) of atomic ensembles. Bell-Bloom magnetometers feature synchronous pumping, whereas FID magnetometers allow direct determination of the magnetic field magnitude from the decay signal frequency \cite{hunter2018free,dong2016high,xu2025free,yang2021all} without prior estimation of the magnetic field. The combination of these two schemes enables both low crosstalk and fast measurement, which serves as the primary motivation for developing the Bell-Bloom-type FID magnetometer in this work.

In this paper, we propose a scheme combining orthogonally polarized counter-propagating pumping and multi-pass probe detection, and experimentally demonstrate a Bell-Bloom-type FID rubidium atomic magnetometer. Counter-propagating pump beams compensate for pump attenuation, homogenize pumping distribution, and suppress axial spin polarization gradients, while an extra-cell five-pass probe scheme extends the light-atom interaction length to enhance FID signal and SNR. A precise pump-probe time-separation strategy eliminates pump-induced effects, including light shifts and power broadening. Comparative experiments between the proposed scheme and the conventional single-beam single-pass configuration show that the magnetometer achieves a sensitivity of 18.9 pT/$\sqrt{\text{Hz}}$ in the traditional mode, while the optimized system reaches $3.1 \ \text{pT}/\sqrt{\text{Hz}}$, corresponding to an approximately sixfold improvement in sensitivity. The proposed scheme not only enhances the magnetic field detection sensitivity, but also effectively improves the measurement accuracy.  This work provides an effective technical route for optimizing the performance of all-optical atomic magnetometers.

\section{ Principle Description}

 The experimental principle of the Bell–Bloom-type FID magnetometer is illustrated in Fig. 1. The intensity of the $x$-directed pump light is modulated at the Larmor frequency, establishing transverse spin polarization through synchronous pumping. The pump light is then switched off, and the atomic magnetic moments undergo spin precession about the $z$-axis under the static field $B_0$. A probe laser incident along the $y$-direction detects the optical rotation of the probe beam induced by the spin precession. The pump and probe beams are temporally separated to suppress light shifts and power broadening effects caused by the pump light. The timing diagram is shown in Fig. 1(b). The pump–probe pulse period is $T$, divided into two phases: preparation of atomic spin polarization ($T_{\text{pump}}$) and detection of FID signals ($T_{\text{probe}}$).

\begin{figure}[htbp]
\centering\includegraphics[width=10.5cm]{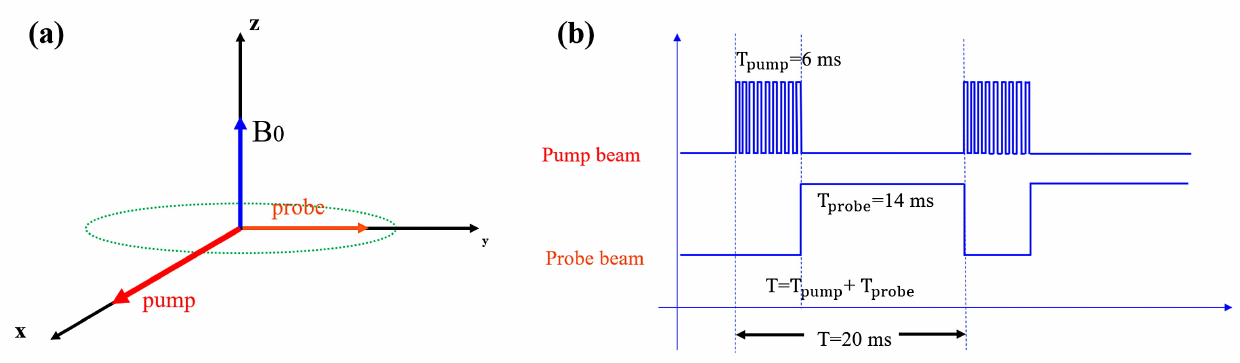}
\caption{(a) Schematic of the Bell-Bloom FID magnetometer in the laboratory frame. The atomic ensemble is placed in a static magnetic field $B_0$ along the z-axis to be measured. Pump light incident along the x-direction spin-polarizes the atoms; probe light incident along the y-direction detects the precession signal of atomic spins in the magnetic field $B_0$; (b) Schematic of pump-probe timing control. During the pump phase of duration $T_{\text{pump}}$, transverse spin polarization is generated. The pump light is then switched off, and the probe light is turned on to monitor the FID signal for a duration $T_{\text{probe}}$.}
\end{figure}

 Under conditions of 100 Torr nitrogen pressure and a temperature well above room temperature, the hyperfine interaction of rubidium atoms is neglected. For the two ground-state sublevels of rubidium atoms ($m_J = -1/2$ and $m_J = 1/2$), $\sigma^+$circularly polarized light resonantly excites atoms from the $m_J = -1/2$ sublevel to the excited state. The excited atoms subsequently decay spontaneously back to the ground-state sublevels. Owing to collisional relaxation induced by the buffer gas, atoms decay into the $m_J = -1/2$ and $m_J = 1/2$ sublevels with equal probability. Notably, atoms residing in the $m_J = 1/2$ sublevel do not interact with the $\sigma^+$ light field. Through repeated cycles of optical excitation and spontaneous decay, most ground-state atoms are eventually pumped into the $m_J = 1/2$ sublevel, generating a macroscopic spin polarization vector parallel to the angular momentum direction of incident $\sigma^+$circularly polarized light.
 
 \begin{figure}[htbp]
\centering\includegraphics[width=8cm]{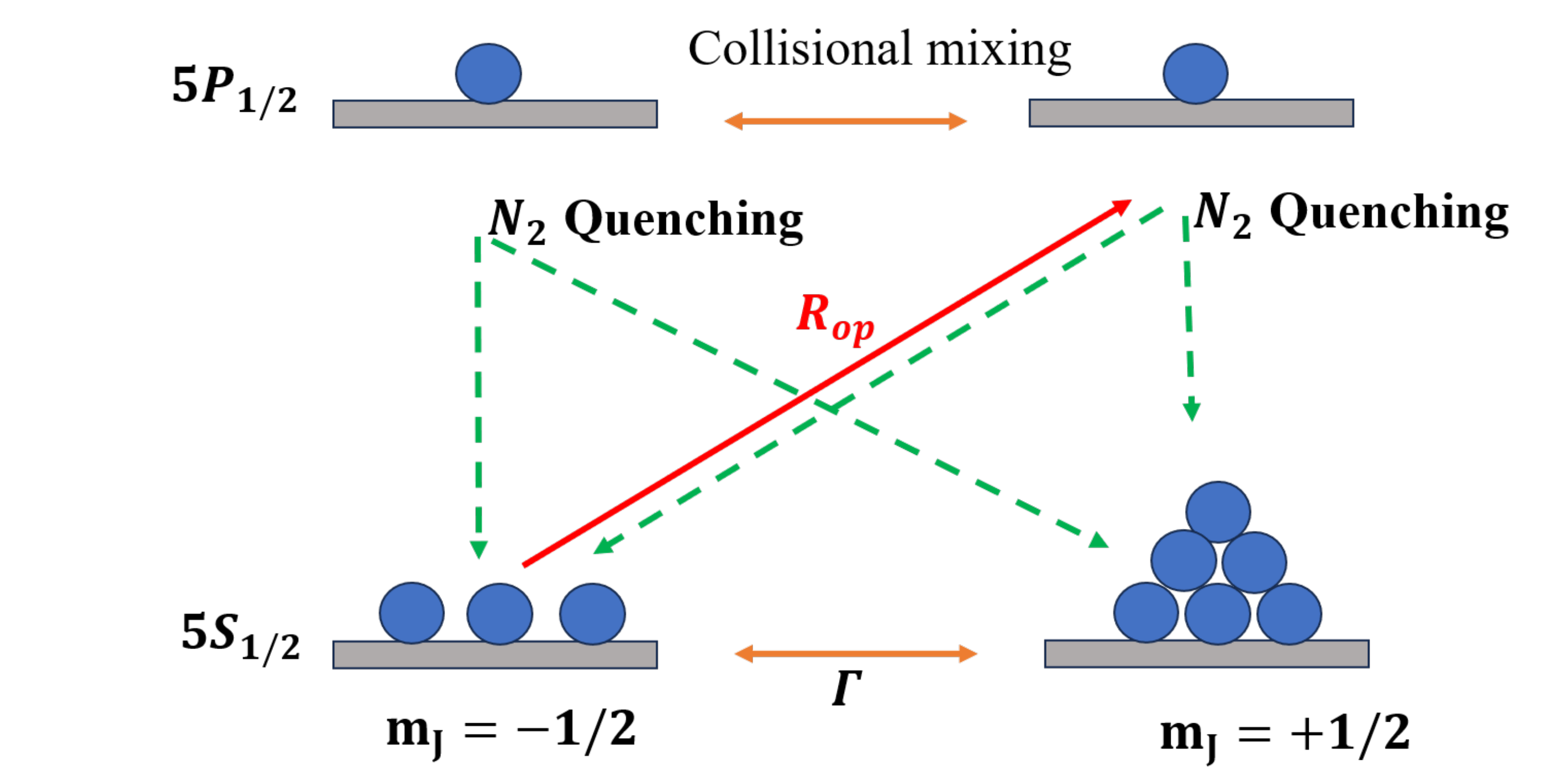}
\caption{Principle of spin polarization}
\end{figure}

 The evolution of the macroscopic atomic spin polarization vector $\mathbf{P}$ in a magnetic field is governed by the Bloch equation:
 
\begin{equation}
    \frac{d\mathbf{P}}{dt} = \gamma \mathbf{B} \times \mathbf{P} + R(s\hat{x} - \mathbf{P}) - R_{\text{rel}}\mathbf{P}
    \label{eq:bloch_original}
\end{equation}

where $\gamma \approx 7 \ \text{Hz/nT}$ is the ground-state gyromagnetic ratio of $^{87}$Rb. $R$ is the optical pumping rate, $s\hat{x}$ is the spin angular momentum of the pump light, and $R_{\text{rel}}$ is the spin relaxation rate. the initial direction of the spin polarization is determined by the pump light, which is considered to be along the x direction in this work.  
During the pump phase, the dominant factor that affects the spin polarization distribution is $R$. Solving Equation (1) in the steady state gives the x-component of the spin polarization vector,

\begin{equation}
    P_x = \frac{R(R + R_{\text{rel}})}{(R + R_{\text{rel}})^2 + (\gamma B_0)^2}
    \label{eq:px_steady}
 \tag{2}
\end{equation}
this relation shows that spin polarization 
$P_x$ is directly proportional to the local optical pumping rate $R$, which means that any spatial variation in $R$ will lead to an inhomogeneous spin polarization distribution.

  As illustrated in Fig. 3, when circularly polarized near-resonant or resonant light propagates through a vapor cell, it is partially absorbed by the alkali metal vapor, causing the pump intensity to attenuate exponentially along the propagation path. This results in an inhomogeneous spatial distribution of $R$ and thus an inhomogeneous spin polarization within the $^{87}$Rb atomic ensemble. Orthogonally polarized $\sigma^+$ and $\sigma^-$ beams are used, collinearly propagating in opposite directions. Their complementary superposition cancels absorption attenuation, homogenizes the pump intensity, and thus suppresses the axial spin polarization gradient. With angular momentum defined along $\pm x$, both beams yield positive projections onto the positive x-axis, delivering identical polarization effects. Their orthogonal polarizations further eliminate interference between the pumps.
  
\begin{figure}[htbp]
\centering\includegraphics[width=10.5cm]{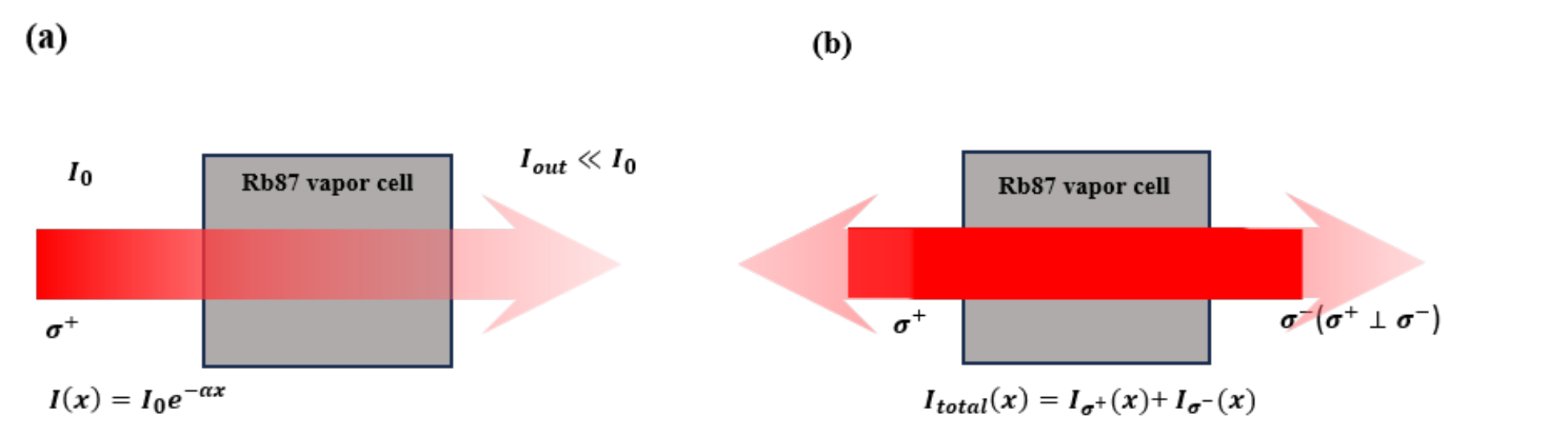}
\caption{Schematic comparison between single‑beam and counter‑propagating orthogonally circularly polarized pumping. (a) In conventional single‑beam pumping, resonant absorption causes exponential intensity attenuation, resulting in a large spin polarization gradient; (b) In the counter‑propagating scheme, orthogonally polarized $\sigma^+$ and $\sigma^-$ beams propagate oppositely and collinearly. Their superposition compensates for absorption, homogenizes the pump intensity, and suppresses the axial spin polarization gradient, improving spatial uniformity.}
\end{figure}

Let the initial intensity of pump beam 1 propagating along the $+x$ direction be $I_{01}$, and that of pump beam 2 propagating  along the $-x$ direction be $I_{02}$. According to the Beer-Lambert law, the intensity of pump light propagating along the $+x$ direction can be expressed as $I_1(x) = I_{01} \exp\left(-n\sigma_a \int_0^x \left[1-P_x(x')\right]dx'\right)$, and the corresponding intensity of pump light propagating along the $-x$ direction can be expressed as $I_2(x) = I_{02} \exp\left(-n\sigma_a \int_x^l \left[1-P_x(x')\right]dx'\right)$, where $n$ is the  atomic number density of the alkali metal, $\sigma_a$ is the resonant absorption cross-section, and $l$ is the length of the vapor cell. $P_x(x)$ denotes the atomic spin polarization at position $x$ within the vapor cell.

The optical pumping rate is linearly related to the pump light intensity, the variation of the pumping rate of the $\sigma^+$  circularly polarized light propagating along the $x$ direction can be expressed as:
\begin{equation}
\frac{dR_1(x)}{dx} = -nR_1(x) \left[1-P_x(x)\right], \quad R_1(0)=R_{01}
\label{eq:pumping_rate1}
\tag{3}
\end{equation}
The variation of the pumping rate of the $\sigma^-$  circularly polarized light propagating along the $-x$ direction can be expressed as:
\begin{equation}
\frac{dR_2(x)}{dx} = nR_2(x) \left[1-P_x(x)\right], \quad R_2(l)=R_{02}
\label{eq:pumping_rate2}
\tag{4}
\end{equation}
 $R_{01}$ and $R_{02}$ represent the initial optical pumping rates at x=0 and x=l, respectively.
 The solutions are as follows

\begin{equation}
R_1(x) = R_{01} \exp\left(- n \int_{0}^{x} \left[1 - P_x(x')\right] \mathrm{d}x' \right)
\label{eq:R1_solution}
\tag{5}
\end{equation}
\begin{equation}
R_2(x) = R_{02} \exp\left(- n \int_{x}^{l} \left[1 - P_x(x')\right] \mathrm{d}x' \right)
\label{eq:R2_solution}
\tag{6}
\end{equation}

When the intensities of the two counter-propagating beams are equal, $R_{01}=R_{02}=R_0$, the total optical pumping rate can be expressed as:
\begin{equation}
R(x) = R_1(x) + R_2(x) = R_{0} (\exp\left(- n \int_{0}^{x} \left[1 - P_x(x')\right] \mathrm{d}x' \right) +  \exp\left(- n \int_{x}^{l} \left[1 - P_x(x')\right] \mathrm{d}x' \right)
\label{eq:total_pumping_rate}
\tag{7}
\end{equation}

 The total pumping rate $R(x)$ exhibits a symmetric distribution along the $x$-direction, which effectively cancels the light intensity gradient induced by single-beam optical pumping and thereby achieves uniform spin polarization of the $^{87}$Rb atomic ensemble.

\section{Experimental Setup}
A schematic of the experimental setup utilized in this work is presented in Fig. 4. The pump beam from a distributed Bragg reflector (DBR) laser sequentially passes through an acousto-optic modulator (AOM) with polarization-maintaining fibers and is expanded to a waist diameter of 6 mm by a beam expander. After traversing a polarizer, the beam is split into two components by a non-polarizing beam splitter (NPBS). Each component is converted into circularly polarized light by a quarter-wave ($\lambda/4$) plate, yielding orthogonal polarizations and perfectly overlapping paths. The two beams then counter-propagate through the atomic vapor cell to achieve atomic spin polarization. The frequency of the pump 
beam is tuned to the optical transition (F = 2 → F’ = 1) of the $^{87}$Rb D1 line, and its frequency is stabilized using a feedback system based on saturated absorption spectroscopy. The probe beam is generated by a second DBR laser at 795 nm. After passing through an AOM and a polarizer, it is prepared as linearly polarized light to detect the atomic spin precession signal. With a spot diameter of 1.4 mm, the probe beam is red-detuned by approximately 40 GHz relative to the $F=1\rightarrow F'=1$ transition of the $^{87}$Rb D1 line, effectively reducing atomic absorption and ensuring detection accuracy.

\begin{figure}[htbp]
\centering\includegraphics[width=10.5cm]{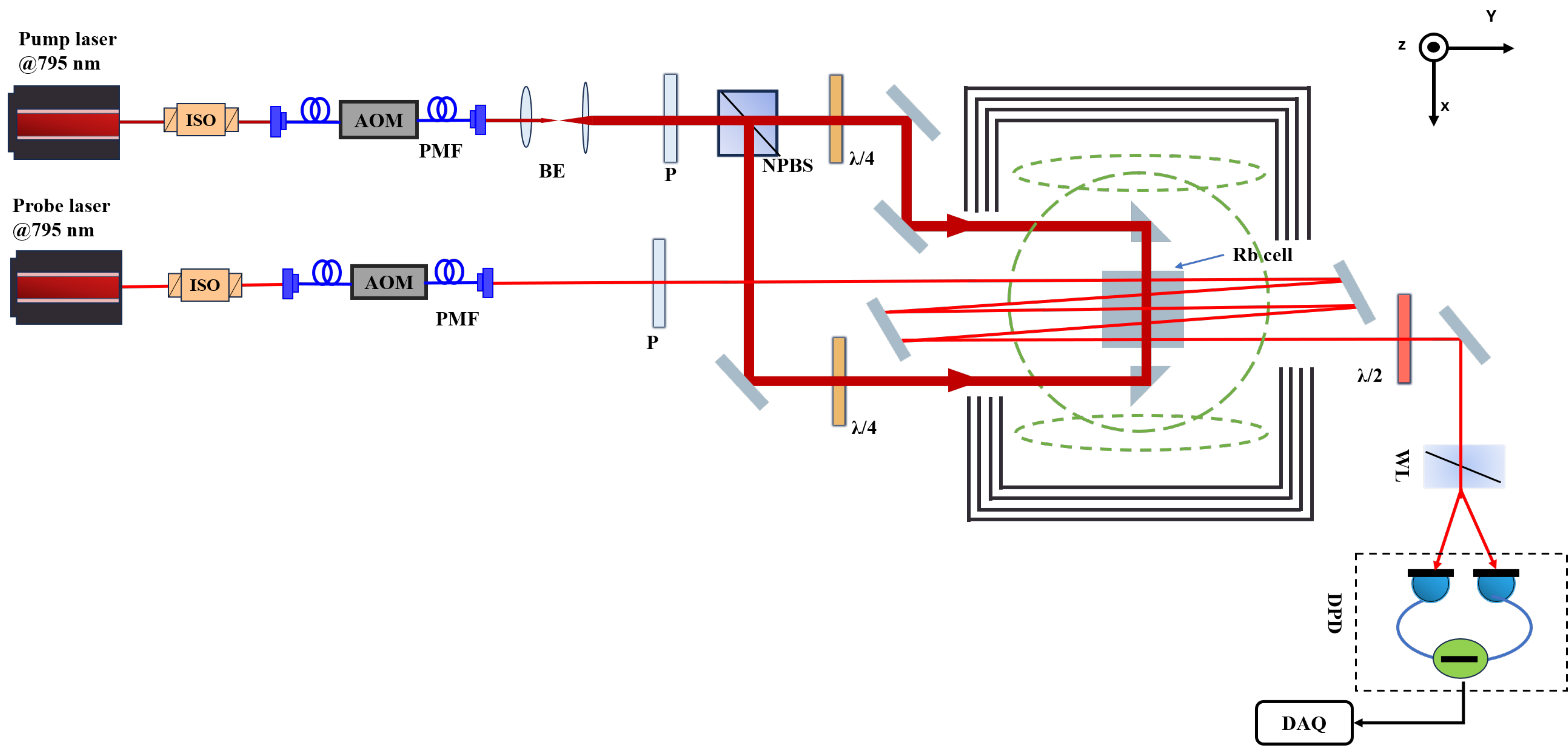}
\caption{Schematic of the experimental setup. ISO: Isolator; AOM: Acousto-optic modulator; PFM: Polarization-maintaining fiber; BE: Beam expander; $\lambda/4$: Quarter-wave plate; $\lambda/2$: Half-wave plate; P: Polarizer; DPD: Differential photodetector; DAQ: Data acquisition card.}
\end{figure}
We use a $15\times15\times15$ mm cubic atomic vapor cell filled with $^{87}$Rb atoms and 100 Torr nitrogen as buffer and fluorescence quenching gas. The cell is placed inside a four-layer $\mu$-metal magnetic shield to suppress ambient magnetic noises, and is uniformly heated by a boron nitride ceramic furnace integrated with flexible thin-film heaters, which are driven by high-frequency alternating current through twisted-pair wiring. A non-magnetic PT100 thermistor is used to monitor the vapor cell temperature. The static bias field $B_0\approx1~\mu$T is generated by a magnetic coil, and the pulse period $T$ is set to 20 ms. During the 6 ms pumping phase, the pump beam is switched and modulated at the Larmor frequency $\omega_L=\gamma B_0$ with a modulation depth of  100\%and a duty cycle of 20\% to fully polarize the atoms. After pumping, the pump beam is blocked and the probe beam is enabled. A balanced polarimeter consisting of a $\lambda/2$ plate, a Wollaston prism, and a balanced detector measures the probe polarization rotation signal over the subsequent 14 ms.

The intensity and effective duration of the light–atom interaction directly determine the amplitude and signal-to-noise ratio (SNR) of the detection signal in atomic magnetometers.  Extending the optical path to enhance the light–atom interaction in the vapor cell is a key strategy to improve detection performance. As shown in Fig. 5, mainstream optical path extension methods include internal multi-reflection cavities and external multi-reflection configurations. The internal scheme achieves stable multi-reflection via in-cell high-reflectivity mirrors \cite{li2022enhancing,sheng2013subfemtotesla,lucivero2021femtotesla} but suffers from high fabrication complexity, alignment difficulty, and cost. In contrast, the external scheme, which extends the optical path through external reflectors without modifying the vapor cell, is easy to implement, flexible, and inexpensive, although its number of reflections is limited by reflector alignment, light loss, and path stability \cite{zhang2022multi}.

\begin{figure}[htbp]
\centering\includegraphics[width=10.5cm]{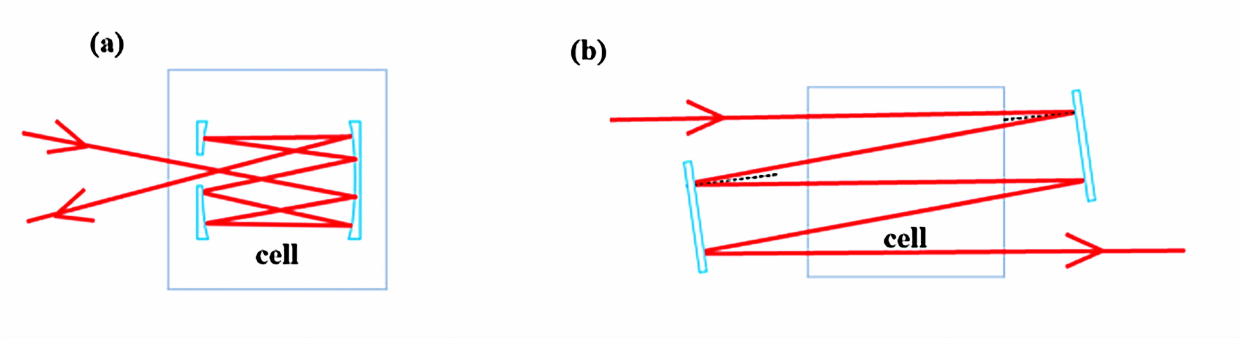}
\caption{Configuration of the multi-pass cell. (a) Multi-reflection cavity inside the gas cell; (b) Multi-reflection outside the gas cell.}
\end{figure}

In this work, we adopt the extra-cell multi-reflection scheme. A pair of high-reflectivity mirrors are placed in parallel outside the atomic vapor cell, with their tilt angles and positions precisely adjusted to enable five stable, low-loss passes of the probe beam through the cell, forming a robust multi-pass detection optical path. This approach avoids the drawbacks of the internal scheme, effectively extends the optical path length, and enhances the light–atom interaction strength, providing reliable support for improving signal amplitude and SNR.

\section{Results and Discussion}
Atomic number density is a critical parameter. A higher atomic number density leads to a larger spin polarization gradient induced by the absorption of pump light, thereby making the compensation and suppression of such a polarization gradient by another orthogonally polarized counter-propagating beam more pronounced. However, with increasing atomic number density, the atomic spin relaxation time decreases rapidly, which significantly degrades the system performance. In the experiment, the temperature of the atomic cells was controlled at $90^\circ$ C, corresponding to a rubidium atomic number density of $3.1\times10^{12} \ \text{cm}^{-3}$. Fig. 6 shows typical FID signals and their fast Fourier transform (FFT) spectra under four different pump-probe configurations: single-beam pumping with single-pass detection, counter-propagating pumping with single-pass detection, single-beam pumping with five-pass detection, and counter-propagating pumping with five-pass detection. All basic parameters are kept constant including the diameters of the pump and probe beams, while only the pumping scheme and the probe optical path length are varied.

\begin{figure}[htbp]
\centering
\includegraphics[width=12.5cm]{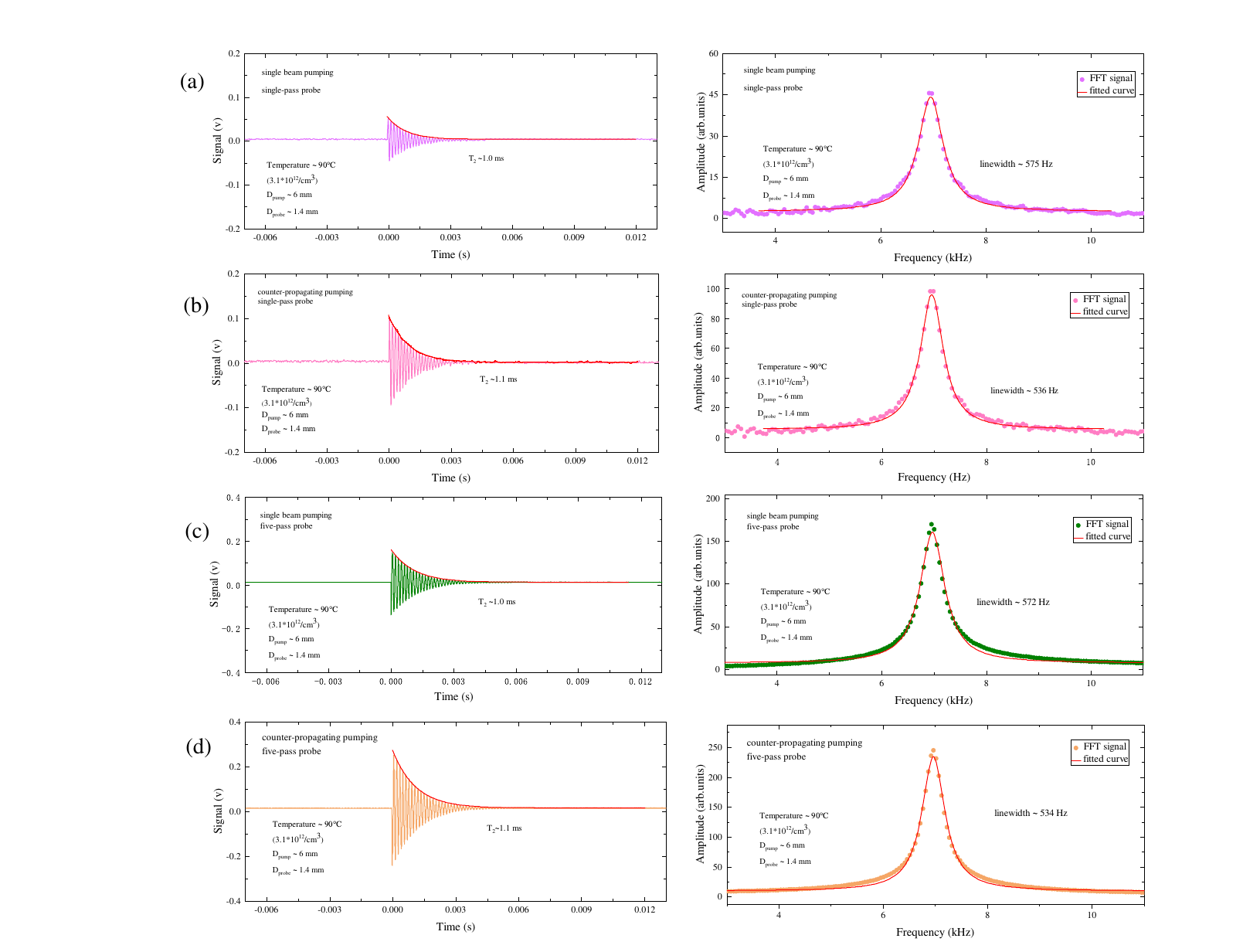} 
\caption{Typical free induction decay (FID) signals and their fast Fourier transform (FFT) spectra under different pump-probe configurations. (a) single-pass probe with single-beam pumping; (b) single-pass probe with counter-propagating pumping; (c) five-pass probe with single-beam pumping; (d) five-pass probe with counter-propagating pumping. Experimental conditions: Temperature: $90^\circ$ C (atomic number density: $3.1\times10^{12}\ \text{cm}^{-3}$), pump beam spot diameter $D_{\text{pump}} \approx 6\ \text{mm}$, probe beam spot diameter $D_{\text{probe}} \approx 1.4\ \text{mm}$.}
\end{figure}

Comparison of Figs. 6(a) and 6(b) shows that for single-pass probe detection, counter-propagating pumping significantly increases the FID signal amplitude and narrows the magnetic resonance linewidth from, indicating improved atomic polarization efficiency and suppressed spin decoherence. Comparison of Figs. 6(a) and 6(c) reveals that under single-beam pumping, five-pass probe detection greatly enhances the FID amplitude while leaving the linewidth nearly unchanged, demonstrating that multi-pass detection boosts the signal via stronger light-atom interaction without altering the dominant linewidth-broadening mechanism. Fig. 6(d) presents the combined configuration of counter-propagating pumping and five-pass detection, which yields the maximum FID amplitude and minimum linewidth, verifying the synergistic optimization of the two schemes.

 The amplitude of the FFT signal is positively correlated with the atomic spin polarization, which is determined by both the magnitude and the spatial uniformity of the optical pumping rate. The pump intensity governs the pumping rate, and its spatial distribution directly affects the spin polarization gradient. To reveal the performance optimization mechanism of the counter-propagating pumping configuration, we systematically measured and compared the FFT signal amplitude and linewidth of the FID resonances under single-beam and counter-propagating pumping with fixed total pump power. As shown in Fig. 7, the signal amplitude under counter-propagating pumping is consistently higher and exhibits similar saturation behavior with increasing pump power. This confirms that counter-propagating pumping alleviates pump absorption attenuation, homogenizes the intensity distribution inside the vapor cell, and effectively suppresses the axial spin polarization gradient, thereby improving the overall atomic spin polarization. Furthermore, the resonance linewidth under counter-propagating pumping is narrower and more stable against pump power variations, which further validates its intensity homogenization effect. The observed power saturation in both schemes originates from optical pumping saturation.

\begin{figure}[htbp]
\centering\includegraphics[width=11.5cm]{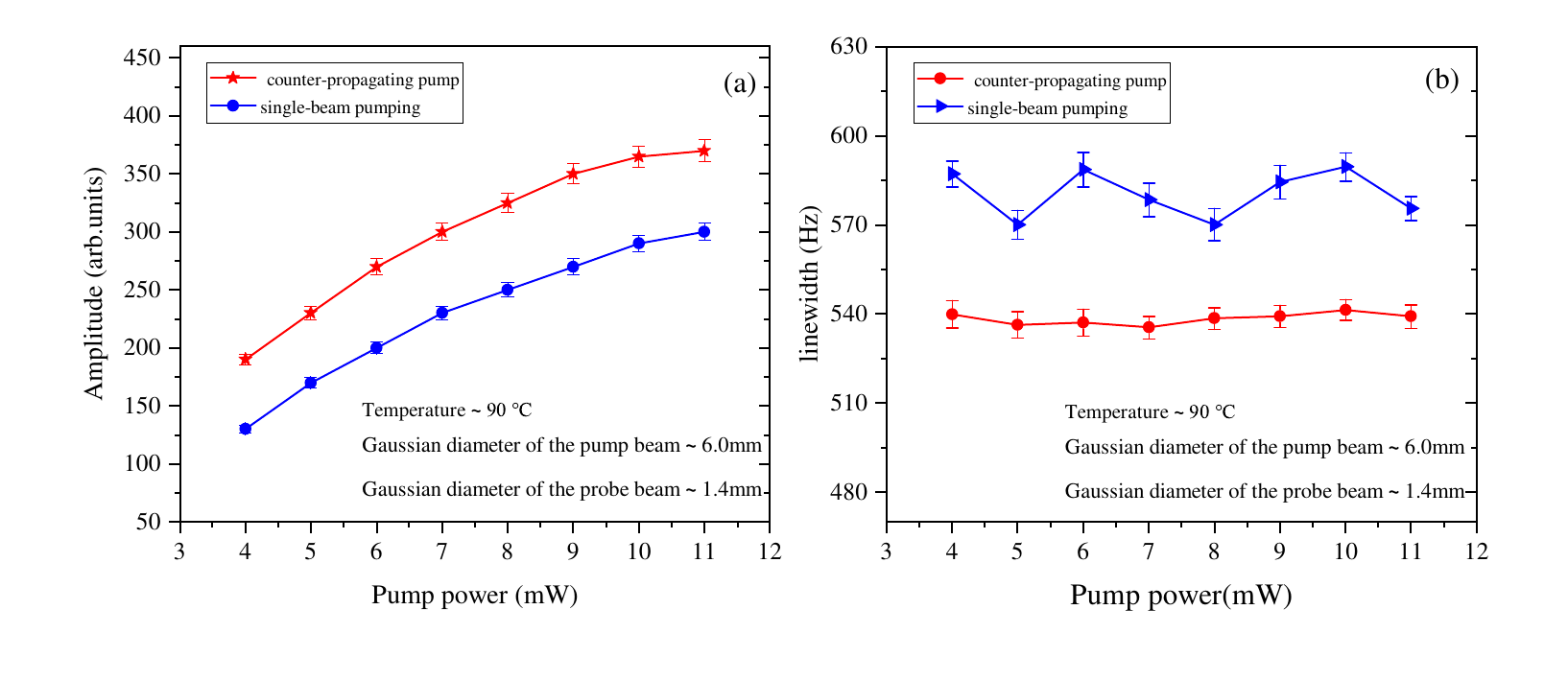}
\caption{Amplitude (a) and linewidth (b) of FFT of FID signals for counter-propagating and single-beam pumping at different pump powers. The total pump power is fixed, with all other conditions unchanged.}
\end{figure}

To quantitatively evaluate the influence of the pumping configuration on magnetic field measurement performance, a high-precision constant current power supply (B2961A) was used to drive a Helmholtz coil and generate the magnetic field $B_0$ under test. We compared the magnetic field measurement accuracy between single-beam and counter-propagating pumping schemes. A data acquisition card with a 500 kHz sampling rate was employed to continuously record magnetic field detection signals, yielding 6000 valid measurement values for each of the four configurations. Gaussian fitting was used to analyze the statistical distribution of the measurements, from which the mean and standard deviation were extracted as key parameters: the mean characterizes the systematic measurement bias, while the standard deviation represents the random measurement error. 

\begin{figure}[htbp]
\centering\includegraphics[width=11.5cm]{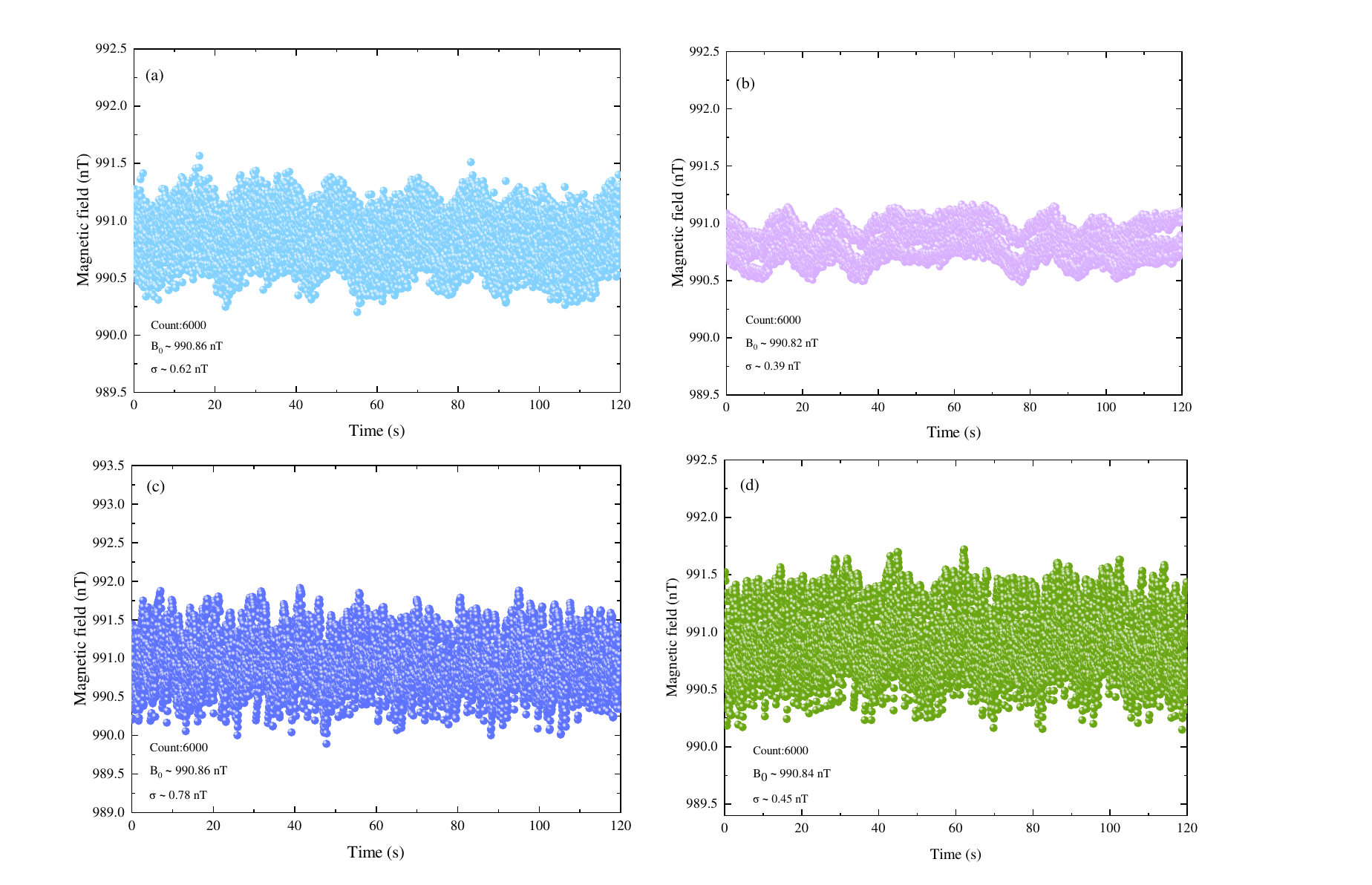}
\caption{Statistical analysis of magnetic field measurements using 6000 data points acquired by a data acquisition card for each configuration: (a) single-beam pumping with five-pass probe detection,  average $B_0$ = 990.86 nT, standard deviation $\sigma$= 0.62 nT;(b) counter-propagating pumping with five-pass probe detection, average $B_0$ = 990.82 nT, standard deviation$\sigma$ = 0.39 nT; (c) single-beam pumping with single-pass probe detection, average $B_0$ = 990.86 nT, standard deviation$\sigma$ = 0.78 nT; (d) counter-propagating pumping with single-pass probe detection, average $B_0$ = 990.84 nT, standard deviation $\sigma$= 0.45 nT.}
\end{figure}

Figure 8 shows that the measured magnetic field means are highly consistent across the four pump-probe configurations. The mean values are 990.86 nT and 990.82 nT for single-beam and counter-propagating pumping with five-pass detection, respectively, and 990.86 nT and 990.84 nT for single-beam and counter-propagating pumping with single-pass detection, respectively. This indicates that the pumping scheme and the number of probe passes through the vapor cell have no significant effect on the systematic bias of the magnetic field measurement.
However, counter-propagating pumping exhibits clear advantages in terms of measurement standard deviation. With five-pass detection, the standard deviation is reduced from 0.62 nT (single-beam) to 0.39 nT (counter-propagating); with single-pass detection, it decreases from 0.78 nT to 0.45 nT. This result demonstrates that counter-propagating pumping compensates for pump-light absorption attenuation along the propagation direction, suppresses local measurement fluctuations induced by the spin polarization gradient, effectively reduces random errors in the magnetic field measurement, and significantly improves the system measurement precision.

In addition, sensitivity is another key performance indicator for atomic magnetometers. Photon shot noise and atomic spin projection noise constitute the fundamental sensitivity limits of atomic magnetometers: photon shot noise arises from statistical fluctuations in the probe photon number, whereas spin projection noise stems from the quantum uncertainty of the atomic spin state. Multi-pass propagation of the probe beam through the vapor cell serves as an effective approach to suppress the contribution of photon shot noise. In this work, the measurement period was set to 20 ms, and the magnetometer bandwidth($f_{\text{bw}}= 1/2T$) was determined as 25 Hz. The magnetic field sensitivity was derived by calculating the power spectral density of the collected magnetic field data and averaging the spectral values over the frequency range of 0.1–25 Hz, with the results presented in Fig. 9. For single-beam pumping, the magnetic field sensitivities are $18.9 \ \text{pT}/\sqrt{\text{Hz}}$ and $9.7 \ \text{pT}/\sqrt{\text{Hz}}$ for single-pass probe detection and five-pass probe detection , respectively. Under the same conditions, the magnetic field sensitivities the corresponding sensitivities are improved to $12.5 \ \text{pT}/\sqrt{\text{Hz}}$ and $3.1 \ \text{pT}/\sqrt{\text{Hz}}$ when counter-propagating pumping. This result indicates that counter-propagating pumping reduces the spin relaxation noise by improving the polarization uniformity, while five-pass detection enhances the light-atom interaction efficiency. The synergistic combination of these two strategies yields the optimal improvement in the system’s noise performance.

\begin{figure}[htbp]
\centering\includegraphics[width=11.5cm]{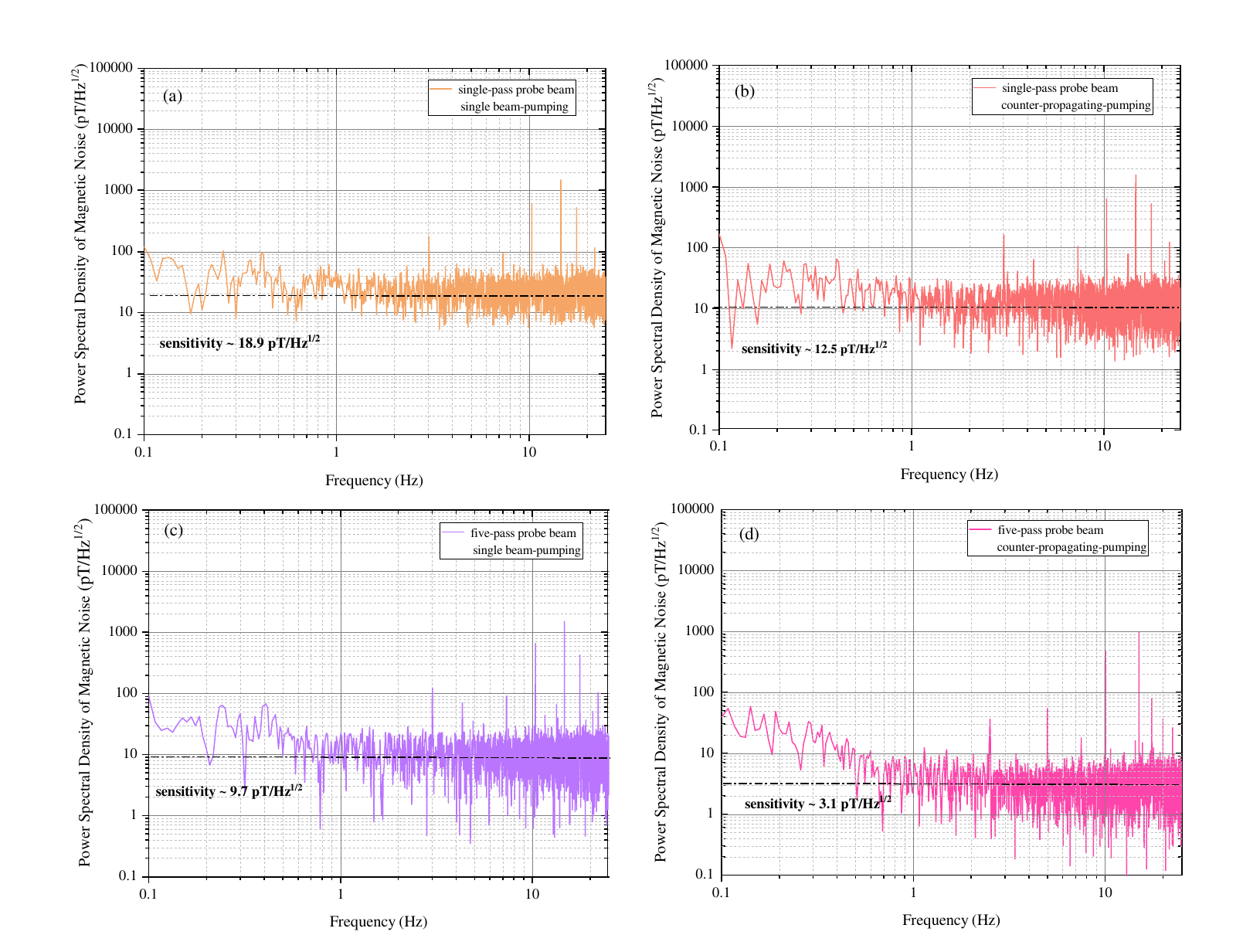}
\caption{Calculated sensitivities of the atomic magnetometer under different pump and probe configurations. The dashed lines represent the estimated noise floor. (a) Single-beam pumping with single-pass probe beam configuration: system sensitivity is approximately $18.9\ \text{pT}/\sqrt{\text{Hz}}$; (b) counter-propagating pumping with single-pass probe beam configuration: system sensitivity is approximately $12.5\ \text{pT}/\sqrt{\text{Hz}}$; (c) Single-beam pumping with five-pass probe beam configuration: system sensitivity is approximately $9.7\ \text{pT}/\sqrt{\text{Hz}}$; (d) counter-propagating pumping with five-pass probe beam configuration: system sensitivity is approximately $3.1\ \text{pT}/\sqrt{\text{Hz}}$.}
\end{figure}

\section{Conclusion}
In conclusion, we have experimentally demonstrated a Bell-Bloom-type FID magnetometer. A counter-propagating pumping scheme with two orthogonally circularly polarized beams is employed to compensate for pump intensity attenuation, thus mitigating the effect of spin polarization gradients on magnetometer accuracy and sensitivity. A five-pass probe configuration is simultaneously utilized to further enhance the SNR. By comparing FID signals between counter-propagating and single-beam pumping, we verify that counter-propagating pumping achieves higher atomic spin polarization and improved measurement accuracy, leading to a sensitivity enhancement from 18.9 pT/$\sqrt{\text{Hz}}$ to 3.1 pT/$\sqrt{\text{Hz}}$.This study is helpful for improving the performance of all-optical atomic magnetometers and is particularly significant for the application of atomic magnetometer arrays. Considering that the intensity of the pump light has a Gaussian distribution, a flat-top beam can be considered as the pump light in the future to construct a more uniform spin-polarized atomic ensemble in three-dimensional space. Additionally, to further break through the standard quantum limit, squeezed light generated via OPO can be introduced\cite{troullinou2021squeezed}.

\begin{backmatter}
\bmsection{Funding}
The National Natural Science Foundation of China (Grant No. 12474483), the Basic Research Program of Shanxi Province (Grant No. 202403021211013), and the Teaching Instrument Development Project of School of Physics and Electronic Engineering, Shanxi University.

\bmsection{Disclosures}
The authors declare no conflicts of interest.

\medskip

\bmsection{Data availability}
Data underlying the results presented in this paper are not publicly available at this time but may be obtained from the corresponding author upon reasonable request.

\end{backmatter}


\bibliography{sample}

\end{document}